\documentclass[12pt,preprint]{aastex}

\usepackage{amsfonts}
\usepackage{amsmath}
\usepackage{amssymb}
\usepackage{wrapfig}
\usepackage{graphicx}
\usepackage{bibentry,natbib}
\usepackage{wasysym}
\usepackage{latexsym}
\usepackage{subfigure}

\newcommand{\half}{{\textstyle{1\over2}}}

\newcommand{\be}{\begin{equation}}
\newcommand{\ee}{\end{equation}}
\newcommand{\bea}{\begin{eqnarray}}
\newcommand{\eea}{\end{eqnarray}}
\newcommand{\nn}{\nonumber}

\newcommand{\lapprox}{\lesssim}

\def\bi{\bf}

\newfont{\myfont}{cmmib10}

\newcommand{\bomega}{\hbox{\myfont \symbol{33} }}

\def\half{{\textstyle{1\over2}}}
\def\bi{\bf}
\def\curl{{\rm curl}\,}
\def\grad{{\rm grad}\,}
\def\div{{\rm div}\,}

\slugcomment{to be published in the Astrophys. J.}

\shorttitle{Obliquely rotating pulsars: current screening}
\shortauthors{Yuen \& Melrose}

\begin{document}

\title{Obliquely rotating pulsars: screening of the inductive electric field}

\author{D. B.  Melrose and Rai Yuen}

\affil{SIfA, School of Physics, the University of Sydney, NSW 2006, Australia}

\begin{abstract}
Pulsar electrodynamics has been built up by taking ingredients from two models, the vacuum-dipole model, which ignores the magnetosphere but includes the inductive electric field due to the obliquely rotating magnetic dipole, and the corotating-magnetosphere model, which neglects the vacuum inductive electric field and assumes a corotating magnetosphere. We argue that the inductive field can be neglected only if it is screened by a current, ${\bi J}_{\rm sc}$, which we calculate for a rigidly rotating magnetosphere.  Screening of the parallel component of the inductive field can be effective, but the perpendicular component  cannot be screened in a pulsar magnetosphere. The incompletely screened inductive electric field has not been included in any model for a pulsar magnetosphere, and taking it into account has important implications. One effect is that it implies that the magnetosphere cannot be corotating, and we suggest that drift relative to corotation offers a natural explanation for the drifting of subpulses. A second effect is that this screening of the parallel inductive electric field must break down in the outer magnetosphere, and this offers a natural explanation for the acceleration of the electrons that produce pulsed gamma-ray emission.
\end{abstract}


\section{Introduction}
Pulsars are obliquely rotating, magnetized neutron stars, with the obliqueness described by the angle, $\alpha$, between the rotation axis and the magnetic dipole axis. The period, $P$, and the period derivative, ${\dot P}$, are known for about 2000 radio pulsars. On a $P$--${\dot P}$ diagram (actually a $\log P$--$\log{\dot P}$ diagram), these fall into three classes: normal pulsars, recycled pulsars with smaller $P,{\dot P}$, and magnetars with larger $P,{\dot P}$. The conventional theory of pulsar electrodynamics is built around two incompatible models, which we refer to as the vacuum-dipole model and the corotating-magnetosphere model. In the vacuum-dipole model the magnetosphere is ignored. Energy and angular momentum are carried away from the pulsar by magnetic dipole radiation at the rotation frequency, $\omega=2\pi/P$. By equating the power radiated to the rate of loss of rotational energy, the model is used to derive the age, $P/2{\dot P}$, and surface magnetic field, $B\sin\alpha\propto(P{\dot P})^{1/2}$, conventionally plotted as straight lines on a $P$--${\dot P}$ diagram. In the corotating-magnetosphere model, the magnetosphere is assumed to be corotating with the star, requiring the presence of the corotation electric field, whose divergence implies the Goldreich-Julian charge density, $\rho_{\rm GJ}$. An additional simplifying assumption is made to reduce the electrodynamics essentially to electrostatics: an explicit assumption that achieves this is that the rotation and magnetic axes are aligned, $\sin\alpha=0$. Alternatively this simplification is achieved through the assumption that the magnetosphere is time-independent in a corotating frame \citep{SAF78}. The energy and angular momentum are assumed to be carried away by a pulsar wind that results from plasma escaping along the open field lines in the polar-cap region, defined by those dipolar field lines that extend beyond the light cylinder radius, $r_{\rm lc}=Pc/2\pi$.  This quasi-electrostatic, quasi-stationary model ties many of the details to the stellar surface, notably space-charge-limited flow, gaps, primary particles, pair formation front and the carousel model for drifting subpulses \citep{Sturrock1971,RudermanSutherland1975,FilippenkoRadhakrishnan1982,OuterGap1986a}. There is no formal justification for the dichotomy between these two models, where one includes time-dependent fields only when the plasma is ignored, and the other neglects the time-dependent fields when the plasma is included. There is a nonzero displacement current in an obliquely corotating magnetosphere, but this is neglected through the assumption \citep{SAF78} that the magnetosphere is time-independent in a corotating frame. This and other criticisms of the existing paradigm  \citep{M04} suggest that the formal basis of the theory needs to reconsidered. Indeed when intrinsically time-dependent electric fields are allowed, through inclusion of  the displacement current in Maxwell's equation, the system is found to be violently unstable to the development of large amplitude oscillations that lead to periodic bursts of pair creation \citep{Letal05,BT07,T10}. Moreover, the existing paradigm is not proving effective as an interpretative or predictive tool. This is especially the case for some recently identified, intrinsically time-dependent pulsar phenomena, notably a link between nulling, mode switching, subpulse drifting and abrupt changes in ${\dot P}$ \citep{Ketal06,Wetal07,LH}. These phenomena seem to require a purely magnetospheric interpretation, and the quasi-electrostatic, quasi-stationary theory tied to the stellar surface has had at best limited success as a basis for their interpretation.


In this paper we consider the implications of  the inductive electric field in an obliquely rotating magnetosphere. The ``inductive'' electric field, ${\bi E}_{\rm ind}$, which has curl ${\bi E}_{\rm ind} \neq 0$ and div ${\bi E}_{\rm ind} = 0$,  results from the time-varying magnetic field, which is an essential ingredient in the vacuum-dipole model. ${\bi E}_{\rm ind}$ is ignored in the corotating-magnetosphere model. This neglect is implicit in the assumption that the magnetosphere is corotating, which implies that the only electric field in the magnetosphere is the corotation electric field, ${\bi E}_{\rm cor}$. However, even if corotating plasma is present, ${\bi E}_{\rm ind}$ is still generated by the changing magnetic field. The neglect of ${\bi E}_{\rm ind}$ is justified only if it is screened by the magnetospheric plasma. We point out that ${\bi E}_{\rm ind}$ cannot be screened by charges. In principle, the associated displacement current, $\varepsilon_0 \partial {\bi E}_{\rm ind} / \partial t$, can be screened by a current, and this might arguably screen the inductive electric field itself. We calculate the screening current density, and discuss whether or not current screening occurs in a pulsar magnetosphere. We conclude that complete screening of ${\bi E}_{\rm ind}$ cannot occur. An unavoidable consequence is that the electric field in the magnetosphere is not equal to ${\bi E}_{\rm cor}$, so that the magnetosphere cannot be corotating. Inclusion of ${\bi E}_{\rm ind}$ in the theory changes the way we need to think about pulsar electrodynamics.

In \S\ref{sect:electric} we calculate the fields in the vacuum-dipole model, first approximating it by a point dipole, and then generalizing to include the surface charge on a star (the Deutsch model). We write down the fields in the corotating-magnetosphere model for an oblique rotator. In \S\ref{sect:screen} we introduce the concept of current screening, and discuss the parallel and perpendicular components of the screening current separately. Perpendicular current screening is argued to be ineffective in a pulsar magnetosphere, and we discuss the resulting departure from corotation in \S\ref{sect:nonrigid}. Possible implications for the interpretation of drifting subpulses and high-energy emission are discussed in \S\ref{sect:discussion}.
 
\section{Electric fields}
\label{sect:electric}

Exact expressions for the pulsar electric field may be written down for three models: the vacuum-dipole model for a point dipole, the Deutsch model for a centered dipole \citep{Deutsch1955}, and an obliquely corotating magnetosphere.

\subsection{Vacuum-dipole model}

The fields from a time-dependent magnetic dipole, ${\bf m}(t)$,  depend on the retarded time $t_{\rm ret}=t-r/c$. The vector potential is
\be
{\bi A}(t,{\bi x})={\mu_0\over4\pi}\,\curl\!\left({{\bi m}(t_{\rm ret})\over r}\right)
={\mu_0\over4\pi}\left[
-{{\bi x}\times{\bi m}\over r^3}
-{{\bi x}\times{\dot{\bi m}}\over r^2c}\right],
\label{Eind1}
\ee
where ${\bf x}$ is the position vector with respect to the center of the pulsar, and where a dot denotes a time derivative. The electric and magnetic fields are determined by
 \be
{\bi E}(t,{\bi x})=-{\partial{\bi A}(t,{\bi x})\over\partial t},
\qquad
{\bi B}(t,{\bi x})=\curl {\bi A}(t,{\bi x}).
\label{Eind2}
\ee
The electric field is given by
 \be
{\bi E}(t,{\bi x})={\mu_0\over4\pi}\left[
{{\bi x}\times{\dot{\bi m}}\over r^3}
+{{\bi x}\times{\ddot{\bi m}}\over r^2c}\right].
\label{Eind3}
\ee 
The magnetic field is given by
\be
{\bi B}(t,{\bi x})={\mu_0\over4\pi}
\left[{3{\bi x}\,{\bi x}\cdot{\bi m}-r^2{\bi m}\over r^5}+{3{\bi x}\,{\bi x}\cdot{\dot{\bi m}}-r^2{\dot{\bi m}}\over r^4c}
+{{\bi x}\times({\bi x}\times{\ddot{\bi m}})\over r^3c^2}
\right].
\label{Eind4}
\ee
We refer to the term in equation (\ref{Eind4}) proportional to ${\bi m}$ as the dipole field, and the terms in equations (\ref{Eind3}) and (\ref{Eind4})  proportional to ${\dot{\bi m}}$ and ${\ddot{\bi m}}$ as the inductive and radiative terms, respectively. In the following discussion, the radiative terms are ignored, except where stated otherwise; for most purposes they can be combined with the inductive terms.

The magnetic field depends on time, with
\be
{\partial{\bi B}(t,{\bi x})\over\partial t}=-\curl{\bi E}(t,{\bi x})
={\mu_0\over4\pi}
\left[{3{\bi x}\,{\bi x}\cdot{\dot{\bi m}}-r^2{\dot{\bi m}}\over r^5}+{3{\bi x}\,{\bi x}\cdot{\ddot{\bi m}}-r^2{\ddot{\bi m}}\over r^4c}
+{{\bi x}\times({\bi x}\times{\dddot{\bi m}})\over r^3c^2}
\right],
\label{Eind5}
\ee
and it has a nonzero curl:
\be
\curl{\bi B}(t,{\bi x})={1\over c^2}{\partial{\bi E}(t,{\bi x})\over\partial t}
={\mu_0\over4\pi c^2}
\left[{{\bi x}\times{\ddot{\bi m}}\over r^3}
+{{\bi x}\times{\dddot{\bi m}}\over r^2c}
\right].
\label{Eind6}
\ee

\subsection{Inductive electric field for a rotating dipole}

For a dipole rotating with angular velocity $\bomega$, one has
\be
{\dot{\bi m}}=\bomega\times{\bi m},
\qquad
{\ddot{\bi m}}=\bomega\times(\bomega\times{\bi m}),
\qquad
{\dddot{\bi m}}=\bomega\times[\bomega\times(\bomega\times{\bi m})].
\label{rf3}
\ee
Of particular interest in the following are the parallel and perpendicular components of the inductive field, where by `parallel' we mean along the dipolar field lines. The parallel component is
\be
E_{{\rm ind}\parallel}(t,{\bi x})=\frac{{\bi E}_{\rm ind}(t,{\bi x})\cdot{\bf B}(t,{\bi x})}{|{\bf B}(t,{\bi x})|}=-{\mu_0\over4\pi r^2}{{\bi x}\cdot{\bi m}\,\bomega\cdot{\bi m}
-{\bi x}\cdot\bomega\,|{\bi m}|^2\over 
[3({\bi x}\cdot{\bi m})^2+r^2|{\bi m}|^2]^{1/2}},
\label{EIp}
\ee
 where `ind' denotes the inductive field. The time derivative of the parallel component of the inductive field is
 \be
{\partial\over\partial t}E_{{\rm ind}\parallel}(t,{\bi x})=-{\mu_0\over4\pi r^2}
{{\bi x}\cdot(\bomega\times{\bi m})\,\bomega\cdot{\bi m}
\over 
[3({\bi x}\cdot{\bi m})^2+r^2|{\bi m}|^2]^{1/2}}.
\label{dEIp}
\ee
The perpendicular component of the inductive field implies a drift velocity given by
\be
\Delta{\bi v}=
{{\bi E}_{\rm ind}(t,{\bi x})\times{\bi B}(t,{\bi x})\over
|{\bi B}(t,{\bi x})|^2}=
{{\bi x}\cdot{\bi m}[2r^2\bomega\times{\bi m}-3{\bi x}\,{\bi x}\cdot(\bomega\times{\bi m})]
\over3({\bi x}\cdot{\bi m})^2+r^2|{\bi m}|^2},
\label{EIperp}
\ee
where only the dipole field is retained in ${\bi B}(t,{\bi x})$.

\subsection{Deutsch model for a centered dipole}

The Deutsch model for a rotating magnetic star has the magnetic field inside a perfectly conducting star described in terms of two functions. For a centered dipole, Deutsch's functions are $R_1(r) ={\mu_0m/2\pi r^3}$ and $R_2(r) ={\mu_0m/4\pi r^3}$. The important change from the point-dipole model is the inclusion of the corotation electric field inside the perfectly conducting star. At the stellar surface ($r=R_*$) with a surrounding vacuum, the tangential component of the electric field must be continuous. This implies a surface charge density on the star, and a potential electric field outside the star. In an aligned model, this potential field is the only electric field present outside the star. In an oblique rotator, the Deutsch model has two contributions to the electric field in the vacuum outside the star:  the potential field (which depends on time) and the inductive field (unchanged from its point-dipole value).

A potential field may be expanded in multipoles, and for a dipolar magnetic field, the potential electric field is quadrupolar. Its explicit form is
\be
{\bi E}_{\rm quad}(t,{\bi x})={\mu_0\over4\pi}{3R_*^2\over5r^7}
\left[
5{\bi x}\,\bomega\cdot{\bi x}\,{\bi m}\cdot{\bi x}
-r^2({\bi x}\,\bomega\cdot{\bi m}+{\bi m}\,\bomega\cdot{\bi x}
+\bomega\,{\bi m}\cdot{\bi x})
\right].
\label{qf6}
\ee
The particular form of ${\bi E}_{\rm quad}$ for an aligned rotator is written down in equation (\ref{eq:InnerPotential}) below. The potential field plays a central role in the conventional model of pulsar electrodynamics. It is assumed to be screened by charges drawn from the stellar surface, with this screening breaking down in gaps,  where the component of ${\bi E}_{\rm quad}$ along the magnetic field lines accelerates charges to high energy. The potential field plays essentially no role in the discussion in the present paper, and it is included here to emphasize that it is unrelated to the inductive electric field.

\subsection{Corotating magnetosphere}

The corotation electric field is determined by the condition that there be no electric field in the local rest frame of the plasma, implying
\be
{\bi E}_{\rm cor}(t,{\bi x})=-(\bomega\times{\bi x})\times{\bi B}(t,{\bi x}).
\label{cf1}
\ee
We emphasize that the assumption of corotation requires that the electric field in the magnetosphere be ${\bi E}_{\rm cor}$, and that if the electric field is not equal to ${\bi E}_{\rm cor}$, the magnetosphere is not corotating. The following remarks relate to implications of the assumption of corotation.

Any rotating vector field, ${\bi V}(t,{\bi x})$, satisfies an equation of motion \citep{Melrose1967}
\bea
{\partial{\bi V}(t,{\bi x})\over\partial t}
&=&\bomega\times{\bi V}(t,{\bi x})-(\bomega\times{\bi x})\cdot\grad {\bi V}(t,{\bi x})
\nn
\\
&=&\curl[(\bomega\times{\bi x})\times {\bi V}(t,{\bi x})]
-(\bomega\times{\bi x})\,\div{\bi V}(t,{\bi x}),
\label{rf2}
\eea
and ${\bi E}_{\rm cor}(t,{\bi x})$ satisfies this equation. All electromagnetic fields in a corotating magnetosphere must satisfy equation (\ref{rf2}).

In an oblique rotator, the divergence, curl and time derivatives of ${\bi E}_{\rm cor}$ are all nonzero.
The divergence of ${\bi E}_{\rm cor}(t,{\bi x})$ determines the corotation charge density:
\be
\div{\bi E}_{\rm cor}(t,{\bi x})=\rho(t,{\bi x})/\varepsilon_0
=-2\bomega\cdot{\bi B}(t,{\bi x})
+(\bomega\times{\bi x})\cdot\curl{\bi B}(t,{\bi x}).
\label{cf2}
\ee
For a dipolar field, equation (\ref{cf2}) reduces to the Goldreich-Julian charge density \citep{Goldreich1969}. The divergence and time derivative (its displacement current) of ${\bi E}_{\rm cor}$ are zero for an aligned rotator, and are specifically neglected by the assumption of time-independent in a corotating frame \citep{SAF78}. 

\subsection{Electric fields in gaps}
\label{sect:MagnetosphericElectricFields}

To emphasize one role of the inductive electric field we discuss conventional models for gaps in an aligned rotator, and comment on how the parallel component of the inductive field requires a change in the interpretation. 

For an aligned rotator, the potential field can be written in the form
\begin{equation} \label{eq:InnerPotential}
	{\bf E}_{\rm quad} = -\nabla\Phi_{\rm quad}, 
	\qquad
	\Phi_{\rm quad} = -\frac{\mu_0}{4\pi}\frac{BR^5\omega}{3r^3}\, P_2({\rm cos}\, \theta),
\end{equation}
where $\Phi_{\rm quad}$ is the potential associated with the surface charge on the star, and where $P_2$ is a Legendre polynomial. This vacuum field has a nonzero parallel component $E_\parallel = {\bf E}_{\rm quad}\cdot{\bf B}/|\bf B|$ along the magnetic field. In a corotating magnetosphere, the electric field given by equation (\ref{eq:InnerPotential}) is perfectly screened, and the only electric field is the corotation field. The concept of a gap is associated with charge starvation, which refers to the situation when there is an insufficient number of charges to provide the Goldreich-Julian charge density. A counterpart of equation (\ref{eq:InnerPotential}) then develops in the gap; the (potential)  electric field in the gap can be attributed to surface charge densities on the lower and upper sides of the gap. Corotation is affected by a gap: the angular velocity of rotation changes across the gap.

In an inner gap model \citep{RudermanSutherland1975}, charges from the stellar surface are assumed to provide the screening charge density immediately above the  surface. In the absence of another source of charge, there is a deficiency in the number of charges needed to maintain $\rho_{\rm GJ}$, and this deficiency increases with height. This leads to $E_\parallel\ne0$ developing in the inner gap; this $E_\parallel$ accelerates  primary particles to high energy in the gap, such that they emit $\gamma$ photons that produce secondary pairs. Charge separation between the secondary pairs provides the additional charge density needed to screen $E_\parallel$ at greater heights. In the original model  \citep{RudermanSutherland1975}, the polar cap region above the gap sub-rotates, i.e., slower than the star to an observer, with $E_\parallel$ proportional to the difference in angular speeds across the gap, and determined by equation (\ref{eq:InnerPotential}) with $\omega$ replaced by this difference and $R$ interpreted as the radial distance to the gap. A criticism of the concept of a quasi-stationary gap is that when time-dependence is included through the displacement current, the gap is violently unstable to the development of large-amplitude electric oscillations, as discussed further below.

Screening can also break down in other regions of the polar cap region, with the outer gap being a notable example. The Goldreich-Julian charge density is zero on surfaces where $\bomega\cdot{\bf B} = 0$, which corresponds to $\cos^2\theta=1/3$, and has opposite signs on either side of this surface along any field line that passes through this surface. An additional source of charge is needed to allow this change in sign to occur. In the absence of an additional source of charge, screening is incomplete and the potential field redevelops, leading to $E_\parallel\neq 0$. As in the inner gap, charges are accelerated and emit gamma rays  \citep{OuterGap1986a}. The location and details of an outer gap model are affected by the current flow through the gap region \citep{Hetal03}. A simple model for the parallel electric field, $E_\parallel$, in the gap is
\begin{equation} \label{eq:OuterE01}
	\varepsilon_0{\partial E_\parallel\over\partial s} = \rho = \rho_{GJ} - e\, [\, N_+({\bf s}) + N_-({\bf s})\, ],
\end{equation}
where $s$ denotes distance along the field line. The term in square bracket is the charge depletion from the Goldreich-Julian value. This electric field  accelerates charges, leading to  pair production. An outer gap is a favored location for the emission of the observed pulsed gamma rays from some pulsars.

The electric fields in gaps in a conventional model are potential fields: they are caused by charges and can be screened by charges. In an oblique rotator, the  presence of ${\bf E}_{\rm ind}$ implies an unrelated contribution to $E_\parallel$, from the parallel component of ${\bi E}_{{\rm ind}}$. This inductive $E_\parallel$ has a different functional form from the potential field; one has $E_{{\rm quad}}\propto\omega/r^4$ and $E_{{\rm ind}}\propto(\omega/r^2)\sin\alpha$. Although relatively unimportant in the inner magnetosphere, $E_{{\rm ind}}$ is likely to be the dominant field in an outer gap region.

\section{Current screening}
\label{sect:screen}

In this section we discuss the concept of current screening in an oblique rotator. The inductive electric field is separated into perpendicular and parallel component, and only the parallel component can be screened by charges. In principle, the perpendicular component can be screened by a current. This current is identified, and whether or not current screening actually occurs is then discussed.

\subsection{Charge screening of the parallel inductive field}

The parallel component of the inductive field can be screened by charges. With $s$ the distance along the field line, one separates the inductive field into parallel and perpendicular components, and writes
\be
\div{\bi E}_{{\rm ind}}={\rm div}_\perp{\bi E}_{{\rm ind}_\perp}+{\partial E_{{\rm ind}\parallel}\over\partial s}=0.
\label{screen1}
\ee
One can then identify a charge density $\rho=-\varepsilon_0\partial E_{{\rm ind}\parallel}/\partial s$. If this charge density is present in the plasma, it produces an $E_\parallel$ that is equal and opposite to $E_{{\rm ind}\parallel}$, effectively screening $E_{{\rm ind}\parallel}$. Provided there is an adequate supply of charge, this screening should occur. Charge starvation can have a similar effect to that in conventional gap models: if there are insufficient charges to screen $E_{{\rm ind}\parallel}$ completely, its presence leads to acceleration of particles, and associated pair production, until there are sufficient charges to restore the screening. 

This argument suggest that the role of ${\bf E}_{\rm ind}$ might be closely analogous to that of the potential field in an aligned model. However, there is a major difference when the displacement current is taken into account. Analytic and numerical solutions of the 1D version of equation (\ref{screen1}) show that it is violently unstable to the build up of large-amplitude  oscillations  in the parallel electric field \citep{Letal05,BT07,T10}. Screening of ${\bf E}_{{\rm ind}\parallel}$ can occur only in a time-averaged sense, where the average is over these oscillations. Pair creation in gaps in the conventional model is replaced by pair creation at phases of the large amplitude oscillations. The concept of a spatially localized gap is no longer relevant. 

\subsection{Displacement current}

The Maxwell equation
\be
\curl{\bi B}(t,{\bi x})=\mu_0{\bi J}(t,{\bi x})+{1\over c^2}{\partial{\bi E}(t,{\bi x})\over\partial t}
\label{cf8}
\ee
must always be satisfied. Current screening corresponds to the displacement current, which is the final term in (\ref{cf8}), being partially or completely replaced by the current ${\bi J}$ carried by the charged particles in the plasma. Before discussing screening, it is relevant to explain how equation (\ref{cf8}) is satisfied in the vacuum-dipole and corotating-magnetosphere models.

For the inductive fields in vacuo, equation (\ref{cf8}) is satisfied with ${\bi J}=0$. The inclusion of a magnetosphere can change all three terms in equation (\ref{cf8}). We assume that the change to $\curl{\bi B}$ is less important than the changes to the two terms on the right hand side. Ideal screening requires  that the final term in equation (\ref{cf8}) be zero. This determines the screening current density as
 \be
{\bi J}_{\rm sc}(t,{\bi x})={1\over\mu_0}\curl{\bi B}(t,{\bi x})=
\varepsilon_0{\partial{\bi E}_{\rm ind}(t,{\bi x})\over\partial t},
\label{scd1}
\ee
where ${\bi E}_{\rm ind}(t,{\bi x})$ is the inductive electric field that would be present in the absence of screening. In identifying the screening current by equation (\ref{scd1}), we effectively require that the displacement current in vacuo be replaced by an identical ${\bi J}$ carried by charges. If the plasma cannot supply this current, then the displacement current is effectively unchanged from its value in vacuo, and ${\bi E}_{\rm ind} (t,{\bi x})$ must have essentially the same value as in vacuo. We argue below that in a pulsar magnetosphere, the parallel component of equation (\ref{scd1}) may be satisfied, but the perpendicular component cannot be satisfied.

For the corotation field, the time-derivative of equation (\ref{cf1}) can be evaluated using equation (\ref{rf2}), giving
\be
{\partial{\bi E}_{\rm cor}(t,{\bi x})\over\partial t}=\curl[(\bomega\times{\bi x})\times{\bi E}_{\rm cor}(t,{\bi x})]
-(\bomega\times{\bi x})\,\div{\bi E}_{\rm cor}(t,{\bi x}).
\label{cf4}
\ee
One can rewrite equation (\ref{cf4}) in the form of equation (\ref{cf8}), and re-interpret it. The interpretation is as a relation between the displacement current associated with ${\bi E}_{\rm cor}(t,{\bi x})$, the curl of the  corotation-induced magnetic field, ${\bi B}_{\rm cor}(t,{\bi x})=(\bomega\times{\bi x})\times{\bi E}_{\rm cor}(t,{\bi x})/c^2$, and the current density, $\rho_{\rm GJ}\bomega\times{\bi x}$, due to the corotating charge density. The corotation-induced magnetic field is smaller than ${\bi B}(t,{\bi x})$ by a factor of order $r^2/r_{\rm lc}^2$, and can be neglected in the inner magnetosphere. Thus equation (\ref{cf8}) is satisfied by the corotation fields alone. This justifies the neglect of the corotation field in identifying the screening current given by equation (\ref{scd1}).

\subsection{Ideal screening in a corotating model}

The assumption that the magnetosphere is corotating has the implication that the inductive field is absent and hence must be screened. With the inductive field given by equation (\ref{Eind3}), the required current density is
 \be
{\bi J}_{\rm sc}(t,{\bi x})=
{{\bi x}\times[\bomega\times(\bomega\times{\bi m})]\over 4\pi c^2r^3}.
\label{scd2}
\ee
The parallel component of the screening current follows from equation (\ref{dEIp}):
 \be
J_{{\rm sc}\parallel}(t,{\bi x})= -{1\over4\pi r^2c^2}
{{\bi x}\cdot(\bomega\times{\bi m})\,\bomega\cdot{\bi m}
\over 
[3({\bi x}\cdot{\bi m})^2+r^2|{\bi m}|^2]^{1/2}}.
\label{scd3}
\ee

We emphasize that the current density given by equation (\ref{scd2}) is required by the hypothesis that the magnetosphere of an oblique rotator is rigidly corotating. In order for this to be the case, the plasma must supply the current ${\bi J}_{\rm sc}$, given by equation (\ref{scd1}).

\subsection{Parallel current screening}

The parallel component of ${\bi E}_{\rm ind}$ can be screened by charges, and the parallel component of ${\partial{\bi E}_{\rm ind}/\partial t}$ can be screened by the current given by equation (\ref{scd3}). Such screening occurs provided that there are sufficient charges available to provide the charge and current densities. The required number of charges can be estimated as follows.

The inductive and corotational fields are proportional to $\omega/r^2$, and they differ in magnitude only by geometric factors. Hence, the charge density required to screen the $E_{{\rm ind}\parallel}$ differs from the Goldreich-Julian value, $\rho_{\rm GJ}$, only by a similar geometric factor. Screening of $E_{{\rm ind}\parallel}$ occurs provided that secondary pair creation results in a multiplicity, $M$, greater than unity \citep{PhysicsOfThePulsarMagnetosphere}. Specifically, if the number densities of electrons and positrons are $n_\pm$, and $\rho_{\rm GJ}=e(n_+-n_-)$ is the number density required for corotation, then the requirement is that the multiplicity, $M=(n_++n_-)/\rho_{\rm GJ}$, be greater than unity. 

The requirement on the parallel current density for screening of ${\partial E_{{\rm ind}\parallel}/\partial t}$ can be estimated by noting that the maximum current density is when the electrons and positrons are flowing in opposite directions at relativistic speeds. This maximum is $e(n_++n_-)c=M\rho_{\rm GJ}c$. Apart from factors of order unity, one has $J_{{\rm sc}\parallel}/M\rho_{\rm GJ}c\approx  r/Mr_{\rm lc}$, which is much less than unity for $r\ll r_{\rm lc}=c/\omega$. It follows that pair creation (in large-amplitude oscillations) provides an adequate source of charges to ensure that parallel current screening occurs in the inner magnetosphere. However, parallel current screening must break down in the outer magnetosphere. 

\subsection{Perpendicular current screening}

Perpendicular current screening involves different physics from parallel current screening. We discuss this from two complementary viewpoints.

The equivalent dielectric tensor of any plasma at very low frequencies may be approximated by a perpendicular component $1 + c^2/v_A^2$ and a parallel component $1 - \omega_p^2/\omega^2$, where the unit terms corresponds to vacuum. Here the Alf$\acute{\rm v}$en speed is $v_A = B/(\mu_0\eta)^{1/2}$, where $\eta$ is the mass density. In a conventional plasma, one has $v_A \ll c$, but in a pulsar plasma, one has $v_A \gg c$. Hence, the perpendicular response of a pulsar plasma is effectively the same as if the plasma were absent. The perpendicular response links the current density, ${\bi J}_\perp$, to the displacement current, $\varepsilon_0 \partial{\bi E}_\perp/\partial t$. It follows that the plasma can supply only a fraction $c^2/v_A^2 \ll 1$ of the current required to screen the displacement current. In contrast, interpreting $-\omega^2$ as a second time derivative, the parallel response corresponds to $\partial J_\parallel/\partial t = \omega_p^2\varepsilon_0 E_\parallel$, which leads to large amplitude electric oscillations in a pulsar plasma \citep{Letal05}.

The perpendicular plasma response (at very low frequencies) can be understood in terms of the so-called polarization drift. Due to $\partial{\bi E}_\perp/\partial t$, a particle with charge $q$ and mass $m$ drifts across the magnetic field lines at a velocity $(m/qB^2) \partial{\bi E}_\perp/\partial t$. For an inductive electric field due to the magnetic field varying at frequency $\omega=2\pi/P$, the polarization drift is smaller than the drift caused by the inductive electric field by a factor of $\omega/\Omega_e$ (see equation (10)), where $\Omega_e=eB/m$ is the cyclotron frequency. Summing over all charges, this leads to a current density ${\bi J}_\perp = (c^2/v_A^2) \varepsilon_0 \partial{\bi E}_\perp/\partial t$, which reproduces the result implied by the plasma response tensor, providing a physical interpretation of this response.

In summary, the parallel component of the displacement current can be screened by a plasma current, which, however, is unstable to large-amplitude oscillations. The perpendicular component of the displacement current is essentially unchanged from its value in vacuo. It follows that the neglect of the perpendicular component of the inductive electric field in models for pulsar electrodynamics is not justifiable. Implications of including the inductive electric field are discussed in the remainder of this paper.

\section{Departure of corotation}
\label{sect:nonrigid}

The neglect of the inductive electric field in an obliquely rotating pulsar is not justified, and its inclusion implies that the magnetosphere cannot be in rigid rotation.

\subsection{Inductively induced drift velocity}

The perpendicular component of the inductive electric field implies an electric drift $\Delta{\bi v}$, determined by equation (\ref{EIperp}).  It is convenient to introduce spherical polar coordinates, $r, \theta,\phi$ defined by the rotation axis. The spherical polar components of the dipolar field are
\be
\left(
\begin{array}{c}
B_r\\
B_\theta\\
B_\phi
\end{array}
\right)={\mu_0m\over4\pi r^3}
\left(
\begin{array}{c}
2[\cos\alpha\cos\theta+\sin\alpha\sin\theta\cos(\phi-\omega t)]\\
\cos\alpha\sin\theta-\sin\alpha\cos\theta\cos(\phi-\omega t)\\
\sin\alpha\sin(\phi-\omega t)
\end{array}
\right),
\label{Bcomp}
\ee
where the initial conditions are chosen such that the magnetic axis is in the plane $\phi=0$ at $t=0$. The angle $\theta_m$, defined by $\cos\theta_m=\cos\alpha\cos\theta+\sin\alpha\sin\theta\cos(\phi-\omega t)$, corresponds to the magnetic colatitude, which varies periodically as the star rotates. At the phases $\phi-\omega t=n\pi$, $n=0,\pm1,\cdots$, the field line is in an azimuthal plane, but at other phases the field has a nonzero azimuthal component ($B_\phi\ne0$).

The drift velocity given by equation (\ref{EIperp}) has spherical polar components
\be
\left(
\begin{array}{c}
\Delta v_r\\
\Delta v_\theta\\
\Delta v_\phi
\end{array}
\right)={\omega r\,\sin\alpha\,\cos\theta_m
\over1+3\cos^2\theta_m}
\left(
\begin{array}{c}
\sin\theta\,\sin(\phi-\omega t)\\
2\cos\theta\,\sin(\phi-\omega t)\\
2\cos(\phi-\omega t)
\end{array}
\right).
\label{Deltav}
\ee
The drift velocity is perpendicular to the magnetic field, implying $\Delta v_rB_r+\Delta v_\theta B_\theta+\Delta v_\phi B_\phi=0$. The drift velocity has its extrema at $\cos(\phi-\omega t)=\pm1$, with values
\be
\Delta v_\pm
={2\omega r\,\sin\alpha\,\cos(\alpha\mp\theta)
\over1+3\cos^2(\alpha\mp\theta)}.
\label{Deltamax}
\ee
The maximum drift speed occurs for  $\cos(\alpha\mp\theta)=1$, and has value $\Delta v_{\rm max}=\half\omega r\sin\alpha$.

\subsection{Small obliquity approximation}
\label{sect:smallalpha}

In  the case $\sin\alpha\ll1$, the effects of obliquity can be treated as a perturbation on the aligned model, as discussed in the Appendix. To zeroth order in $\sin\alpha$, the magnetosphere is assumed to be rigidly corotating, with the charge density equal to $\rho_{\rm GJ}$. The inductive electric field is of first order in $\sin\alpha$, 
\be
\left(
\begin{array}{c}
E_{{\rm ind}r}\\
E_{{\rm ind}\theta}\\
E_{{\rm ind}\phi}
\end{array}
\right)={\mu_0m\omega\sin\alpha\over4\pi r^2}
\left(
\begin{array}{c}
0\\
-\cos(\phi-\omega t)\\
\cos\theta\,\sin(\phi-\omega t)
\end{array}
\right),
\label{Eind}
\ee
and is unchanged to lowest order in the iteration.

\begin{figure}
\begin{center}
\includegraphics{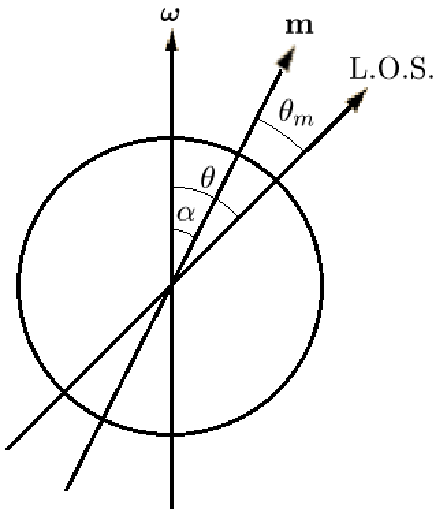}
\caption{The polar angles used in this paper are shown in the plane that contains the rotation and magnetic axes. Angle $\theta$ is the polar angle relative to the rotation axis, and we denote the magnetic colatitude by $\theta_m$. This differs from a notation used in the pulsar literature \citep{LyneManchester1988} where these two angles are denoted $\zeta$ and $\theta$, respectively. The angle $\beta=\theta-\theta_m$ is referred to as the impact parameter. For ${\rm sin}\,\alpha\ll 1$, we have $\alpha\approx\theta\approx\theta_m$.}
\label{fig-GeoImpParam}
\end{center}
\end{figure}

On expanding the drift velocity given by equation (\ref{Deltav}) in $\sin\alpha$, the first order terms average to zero over a rotation period, and we retain a term of second order in $\sin\alpha$ to show that this is not the case in general. The resulting expression for  the drift velocity is
\be
\left(
\begin{array}{c}
\Delta v_r\\
\Delta v_\theta\\
\Delta v_\phi
\end{array}
\right)={\omega r\,\sin\alpha\,\cos\theta
\over1+3\cos^2\theta}
\left[1+{1-3\cos^2\theta\over1+3\cos^2\theta}\sin\alpha\sin\theta\cos(\phi-\omega t)\right]
\left(
\begin{array}{c}
\sin\theta\,\sin(\phi-\omega t)\\
2\cos\theta\,\sin(\phi-\omega t)\\
2\cos(\phi-\omega t)
\end{array}
\right),
\label{Deltav1}
\ee
with the expression inside the square brackets replaced by unity to lowest order. For ${\rm sin}\, \alpha\ll 1$, we have ${\rm cos}\, \theta_m = {\rm cos}(\alpha\mp\theta)\approx{\rm cos}\, \theta$ (see figure \ref{fig-GeoImpParam}). The average drift over a rotation period is in the $\phi$ direction and is of order $\sin^2\alpha$:
\be
\langle\Delta v_\phi\rangle=\zeta\omega r\sin\theta,
\qquad
\zeta=\sin^2\alpha
{\cos\theta\,(1-3\cos^2\theta)\over(1+3\cos^2\theta)^2}.
\label{avvphi}
\ee
This average drift vanishes in the equatorial plane, and has opposite signs in opposite (rotational) hemispheres.

The drift given by equation (\ref{Deltav1}) is superimposed on the assumed rigid corotation. The drift does not correspond to a modification of the angular velocity. To see this consider a perturbation, $\delta\omega$, in the angular speed of the magnetosphere. This would give a perturbation $\delta \omega\, r\sin\theta$ in azimuthal velocity; this has a maximum in the equatorial plane, has the same sign in both hemispheres and is independent of rotational phase. The velocity given by equation (\ref{Deltav1}) satisfies none of these conditions. The magnetosphere cannot be in rigid rotation at any angular velocity.

\begin{figure}[htb]
\begin{center}
\includegraphics{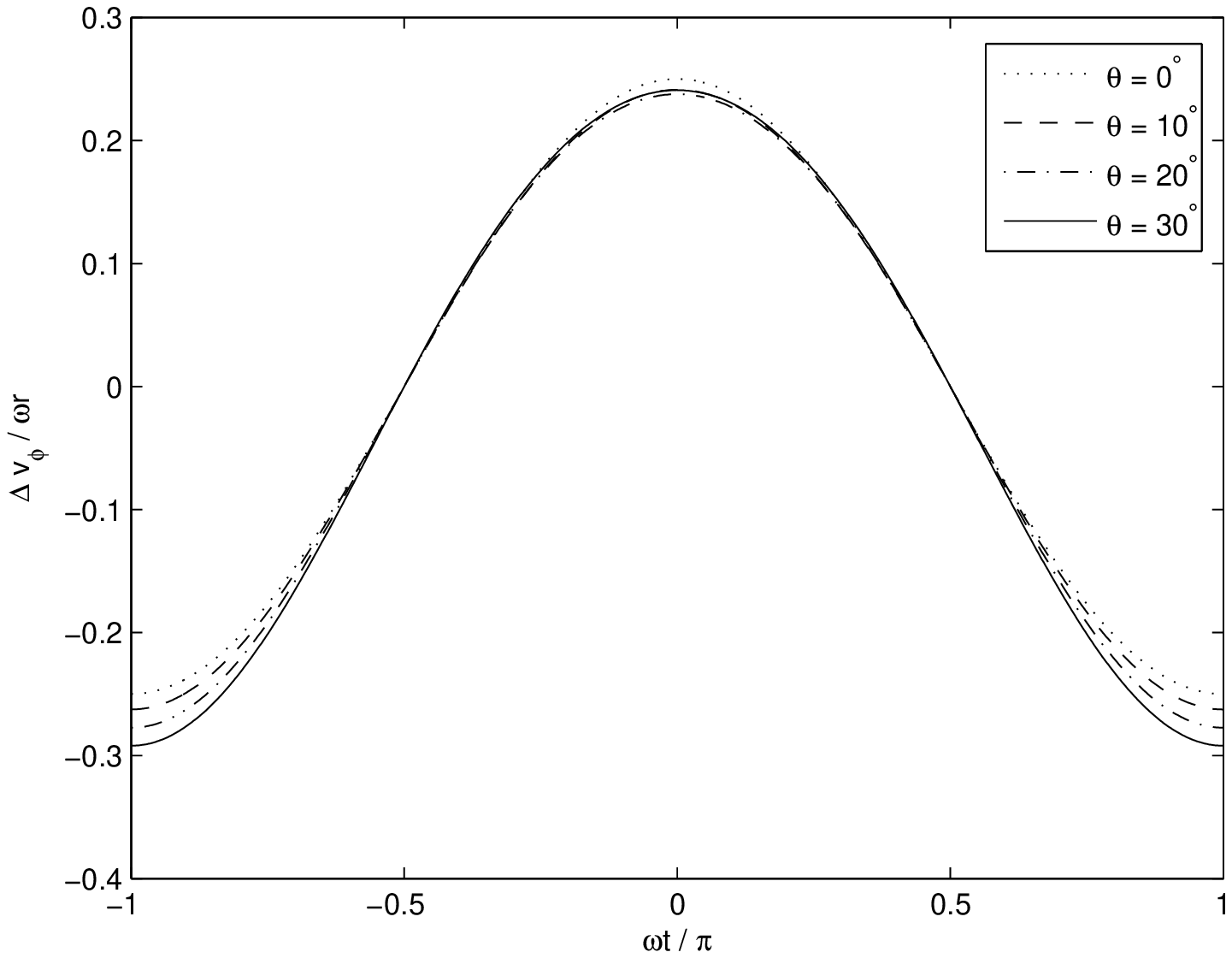}
\caption{The azimuthal drift velocity, $\Delta v_\phi(t,\theta)$, as given by equation (\ref{Deltav1}), plotted against time or rotational phase for $\alpha = 30^\circ$. The different curves represent different $\theta$ values range from $0^\circ - \,30^\circ$. In addition to varying sinusoidally in magnitude, $\Delta v_\phi(t,\theta)$ also varies asymmetrically with the positive maximum value different from the negative minimum value, which indicates that the drift velocity is faster in the second half of the rotation.}
\label{fig-SC28v_ph}
\end{center}
\end{figure}

\begin{figure}[htb]
\begin{center}
\includegraphics{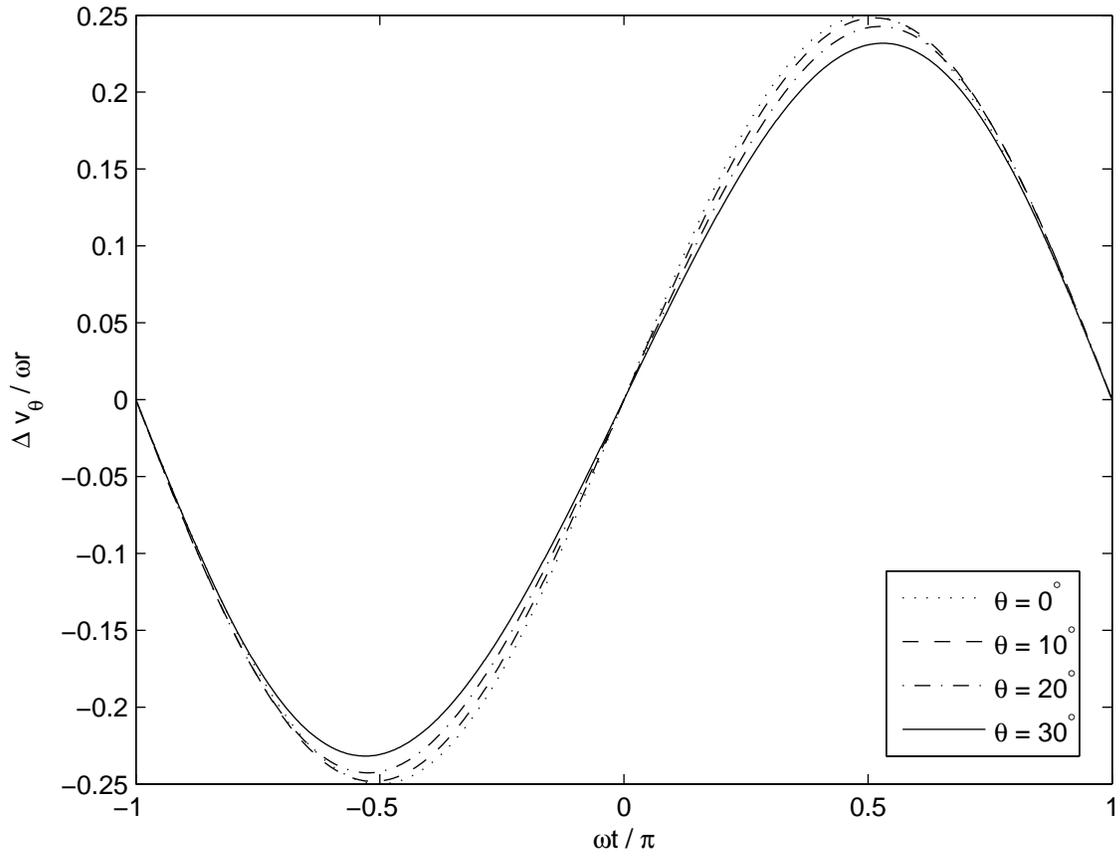}
\caption{As for figure \ref{fig-SC28v_ph} but for the polar component, $\Delta v_\theta(t,\theta)$, as given by equation (\ref{Deltav1}). The amplitude of a curve decreases as $\theta$ increases as oppose to the other two components.}
\label{fig-SC28v_th}
\end{center}
\end{figure}

\begin{figure}[htb]
\begin{center}
\includegraphics{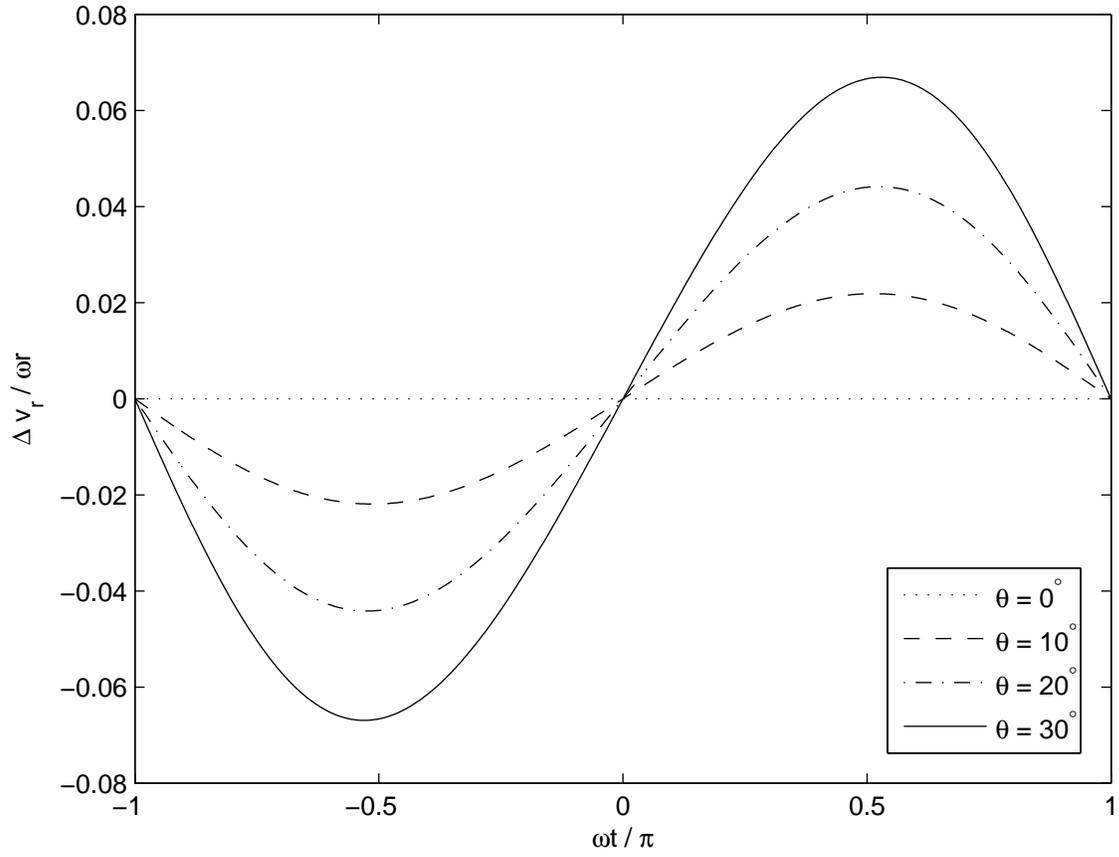}
\caption{As for figure \ref{fig-SC28v_ph} but for the radial component, $\Delta v_r(t,\theta)$, as given by equation (\ref{Deltav1})}.
\label{fig-SC28v_r}
\end{center}
\end{figure}

Figures \ref{fig-SC28v_ph}, \ref{fig-SC28v_th} and \ref{fig-SC28v_r} show plots of $\Delta v_\phi$, $\Delta v_\theta$ and  $\Delta v_r$, respectively, in units of $\omega r$ as functions of $\omega t$ with $t = 0$ at $\phi = 0$. Each curve represents values for $\theta = 0^\circ$, $10^\circ$, $20^\circ$ or $30^\circ$ for $\alpha = 30^\circ$. The correction in square brackets in equation (\ref{Deltav1}) is included, so that the temporal variations include a term varying sinusoidally as $\phi - \omega t$ with a correction term that includes a variation as $2(\phi - \omega t)$. The components $\Delta v_\phi$ and $\Delta v_r$ are odd functions of ${\rm cos}\, \theta$, implying that they have opposite signs in opposite hemispheres, whereas $\Delta v_\theta$ is an even function of ${\rm cos}\, \theta$.

\begin{figure}[htb]
\begin{center}
\includegraphics{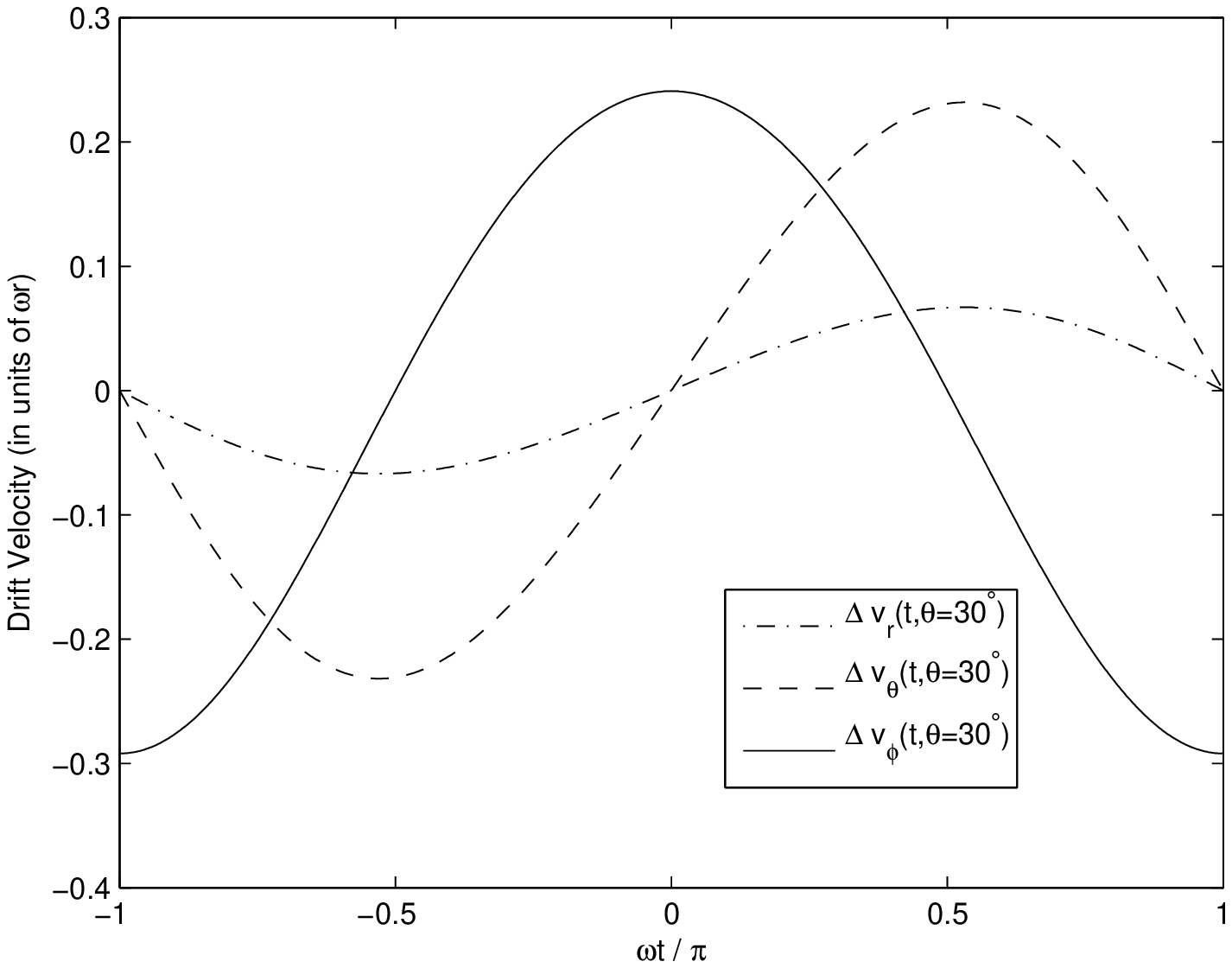}
\caption{The three components $\Delta v_r$, $\Delta v_\theta$ and $\Delta v_\phi$ are shown in the case when the line of sight coincides with the magnetic axis at $\theta = 30^\circ$}.
\label{fig-AsymmetryAlign}
\end{center}
\end{figure}

The pulse window corresponds to a small range of $\phi$ and $\theta$, and because an observer can see the drift only within this window, only a correspondingly small range in figures \ref{fig-SC28v_ph}, \ref{fig-SC28v_th} and \ref{fig-SC28v_r} is relevant to observations. This range is model-dependent, and in most models it corresponds to a range $\Delta\phi$ about $\phi = 0^\circ$ and a range $\Delta\theta$, of order $\Delta\phi$, about a line of sight, $\theta$, close to $\alpha$. The relevant regions in figures \ref{fig-SC28v_ph}, \ref{fig-SC28v_th} and \ref{fig-SC28v_r} are near $\omega t = 0$ for the solid curves. The three curves for $\theta = \alpha = 30^\circ$ are replotted in figure \ref{fig-AsymmetryAlign}. As the variation with $\theta$ is relatively small, the observable (relative to corotation) drifts correspond to the range $\Delta\phi$ about $\omega t = 0$ in figure \ref{fig-AsymmetryAlign}. The components $\Delta v_\theta$ and  $\Delta v_r$ are zero at $\omega t = 0$, implying that the observable drift is in the $\phi$ direction. The magnitude of this drift may be estimated by setting $\theta = \alpha$, $\phi - \omega t = 0$ in equation (\ref{Deltav1}). This implies an observable drift

\begin{equation}
(\Delta v_\phi)_{\rm obs} = \frac{2\omega r \,{\rm sin} \,\alpha \,{\rm cos} \,\alpha}{1 + 3 \,{\rm cos}^2 \,\alpha} \,\Bigg[\, 1 + \frac{1 - 3 \,{\rm cos}^2 \,\alpha}{1 + 3 \,{\rm cos}^2 \,\alpha} \,{\rm sin}^2 \,\alpha \,\Bigg].
\label{ObserveDrift}
\end{equation}

Note that the sign of $(\Delta v_\phi)_{\rm obs}$ depends on the sign of ${\rm cos} \,\alpha$, that is, on the sign of $\bomega\cdot{\bf B}$. This sign also determines the sign of the Goldreich-Julian charge density. The inductively induced drift implies super-rotation when the Goldreich-Julian density corresponds to an excess of electrons, and to sub-rotation when the Goldreich-Julian density corresponds to an excess of positrons.

\section{Discussion}
\label{sect:discussion}

The inductive electric field, ${\bi E}_{\rm ind}$, (due to the obliquely rotating dipole) cannot be completely screened in a pulsar magnetosphere.  Its presence implies a plasma drift across the field lines, such that the motion of the magnetosphere is not rigid corotation with the star. We explore the suggestion that this drift might provide a natural explanation for subpulse drifting. We also comment on the possible role of the inductive field in an outer-gap model for pulsed gamma-ray emission.

\subsection{Subpulse drifting}

The inductively-induced drift relative to corotation suggests a natural explanation for subpulse drifting. Drifting subpulses correspond to a systematic motion of the plasma from which the radio emission escapes within the pulse window corresponding to a small range of phases during which emission is observed. Data on subpulse drifting in a large sample of pulsars \citep{Wetal06,Wetal07} lead to several general properties. One is that the subpulse drift rate is independent of frequency. This is explained naturally by this drifting model, which is essentially geometric and varies only slowly with height, and hence with frequency assuming frequency-to-radius mapping. Another property is that the number of pulsars with positive and negative drift rates are roughly equal. This has a natural explanation due to the drift rate depending on the sign of $\cos\theta$. For small impact parameter, implying $\theta\approx\alpha$ in the polar cap region, the drift rate has one sign for pulsars with $\alpha<\pi/2$ and the opposite sign for pulsars with $\alpha>\pi/2$. The sign of the drift rate depends on the sign of the Goldreich-Julian density above the polar caps.

A simple prediction of the model is that the subpulse drift rate should be proportional to $\sin\alpha$.  This follows from the fact that the drift rate is proportional to $\sin\alpha$ for small $\sin\alpha$. 

A more subtle prediction is based on the sign of the drift rate depending on the sign of $\cos\theta$, which is interpreted as the sign of $\cos\alpha$ for emission from the polar cap region. The sign of $\cos\alpha$ determines the sign of $\rho_{\rm GJ}$. Some pulsars  have a  preferred handedness for circular polarization, and this is plausibly related to the sign of $\rho_{\rm GJ}$ \citep{ML04}. The prediction is that the sign of the subpulse drift rate and the handedness of the net circular polarization should be correlated.

\subsection{Acceleration by the inductive electric field}

Pulsed high-energy emission from pulsars is attributed to acceleration of particles in the outer magnetosphere, for example, due to the breakdown of screening in an outer gap \citep{Tetal10}. In such models, the parallel electric field in the outer gap is assumed to be a potential field. The inductive electric field, which is not included in any existing model, can play the same role as the assumed potential field, suggesting a natural alternative to  outer-gap models. Although the inductive field calculated in section \ref{sect:electric} becomes invalid as the light cylinder is approached, it can be used to provide a rough estimate of the actual electric field. This estimate is $E_\parallel\approx\omega B_*R_*^3/r^2\approx B_{\rm lc}c(r_{\rm lc}/r)^2$, where $B_{\rm lc} = B_\star(R_\star/r_{lc})^3$ is the magnetic field at the light cylinder. The main difference between such an inductive model and the conventional outer-gap models is that the parallel electric field results from the time-varying (obliquely rotating) magnetic field, and is not due to a local charge separation. The parallel inductive electric field appears when there are too few charges to provide the charge density needed to screen it.

There is a close analogy between screening of the inductive field in the magnetosphere ($r<r_{\rm lc}$) and screening of the radiative field in the wind zone ($r\gg r_{\rm lc}$). Early in the development of pulsar theory it was recognized that the radiative component of the vacuum field, given by equations (\ref{Eind3}) and (\ref{Eind4}), would accelerate test particles to extremely high energy \citep{GO69}. In the theory of the pulsar wind, developed subsequently, the electric field is specified by the wind equations, and it is implicit that the radiative field is perfectly screened. This screening, and the generation of the electric field in the wind model require a current in the wind. If there are too few charges (charge or current starvation) in the wind to sustain this current \citep{U94}, the vacuum field is incompletely screened and accelerates charges. The breakdown of perfect screening leads to a transfer of energy from the Poynting flux to a kinetic energy flux, providing a natural explanation for the dominance of the latter in the outer wind zone \citep{MM96}. Similarly, in the inner magnetosphere, an unscreened inductive field accelerates a test charge along a magnetic field line to very high energy. Breakdown of parallel current screening is expected in the outer magnetosphere ($r\lapprox r_{\rm lc}$). The parallel component of the inductive electric field then reappears, and accelerates charges. This needs to be examined in detail as an alternative to an outer gap for the production of pulsed $\gamma$ rays \citep{Hetal03,Tetal10}.

In a conventional pulsar wind model \citep{ReesGunn1974, Arons2004PWN} the magnetic field well beyond the light cylinder has a predominantly toroidal component. The rotation of the star and the outflow of the wind cause the direction of the toroidal component to reverse periodically with radial distance \citep{KirkLyubarskyPetri07}. Specific models \citep{Bogovalov1999, Contopoulos2005} imply a temporally changing electric field in the wind. In such models the fluid theory determines the electric field. Although this electric field is varying periodically with time in an oblique rotator \citep{Bogovalov1999}, implying a nonzero displacement current, this is not the displacement current associated with the rotating dipole. The latter field would be present in vacuo, and its implicitly assumed absence in wind models requires that it be effectively screened by currents. As with the inductive field in the inner magnetosphere, if current screening is ineffective, the radiative electric field must be present in the wind zone.

\section{Conclusions}
\label{sect:conclusions}

The neglect of the inductive electric field, ${\bi E}_{\rm ind}$, in models for a pulsar magnetosphere cannot be justified.  This field is generated by the changing magnetic field and the associated displacement current of an obliquely rotating magnetic dipole. It is impossible in principle to screen an inductive field by charges. We show that screening of the displacement current is possible in principle, and we calculate the required screening current density in an idealized model. Screening of the components parallel to the magnetic field and perpendicular to the field involves different physics. The parallel component is unstable to the development of oscillations, and screening can occur only in an average sense, where the average is over the oscillations. Perpendicular screening involves a polarization current, driven by the displacement current, and can never be complete. Nearly complete screening is possible for $v_A^2\// c^2\ll1$, which is not the case in a pulsar magnetosphere, where the perpendicular inductive field is essentially unscreened, and unchanged from its vacuum value. 

The presence of ${\bi E}_{\rm ind}$ has important consequences for understanding pulsar magnetospheres. In particular, the magnetospheric plasma cannot be corotating with the star. The assumption of corotation is central to conventional corotating-magnetosphere model for pulsars, and much of the physical interpretation of pulsar phenomena is based on this assumption. The inductive electric field is absent only in the artificial case of an aligned rotator ($\sin\alpha=0$). In the case of small obliquity, $\sin\alpha\ll1$, an aligned rotator can be regarded as a zeroth order model, with the inductive field and its implications treated to first order in $\sin\alpha$. Using such a model we calculate the inductively induced drift rate, which is the difference between the actual velocity of the local magnetospheric plasma and the corotation velocity. We suggest that this drift  offers a natural explanation of drifting subpulses. The drifting of subpulses is a direct result of the perpendicular component of the inductive electric field. This suggests that subpulse drift should be interpreted as a signature of the inductive electric field and used to infer properties of this field. 

The parallel component of the inductive field can be screened by charges, and the breakdown of this screening (due to charge starvation) in the outer magnetosphere offers an alternative to conventional (outer gap or slot gap) models for pulsed gamma ray emission. Conventional models invoke $E_\parallel$ due to separation of charges. The conceptual change suggested here is that the (parallel) inductive field would be present in the absence of the screening, and it must appear whenever charge or current starvation limits the ability of the plasma to screen it.

A surprising implication of including the inductive electric field is that the magnetosphere cannot be corotating with the star. The conventional argument is that MHD requires that the electric field be zero in the rest frame of the plasma, and hence is equal to the ${\bi E}_{\rm cor}$ in the pulsar frame. Inclusion of the inductive electric field implies an inductively driven drift, and this drift motion is implicitly and incorrectly assumed to be zero in a corotation model.

Finally, we endorse a remark  that the neglect of the displacement current in astrophysical and space plasmas has led to conceptual misunderstandings \citep{SL06}.  In the context of pulsars, the conventional quasi-electrostatic corotating-magnetosphere model excludes the displacement current, but when the displacement current is included, the system is found to be violently unstable to the development of large amplitude electric oscillations  \citep{Letal05,BT07,T10}.

\section*{Acknowledgments}
We thank Patrick Weltevrede for providing useful comments, Simon Johnston, George Hobbs and Dick Manchester for helpful discussion and advice, and Mike Wheatland for helpful comments on the manuscript.


\appendix
\section{Self-consistent fields}

Both the vacuum-dipole  and the corotating-magnetosphere models are based on assumptions that are not satisfied. The vacuum-dipole model neglects the plasma, but the plasma screens the parallel components of the inductive electric field and the displacement current, invalidating their derivation using the model. The corotating-magnetosphere model is based on the hypothesis that the magnetosphere is corotating, but the perpendicular component of the inductive electric field is essentially unscreened which implies an electric drift incompatible with corotation. This raises the question as to how, in principle,   one can formulate a self-consistent model for the electrodynamics of an obliquely rotating pulsar magnetosphere. In this appendix, some remarks are made on how this might be achieved.

The determination of the fields and induced charge and current densities requires solving a self-consistency problem. The potentials due to given charge and current densities satisfy
\be
\left(
\begin{array}{c}
\phi(t,{\bi x})\\
{\bi A}(t,{\bi x})
\end{array}
\right)={1\over4\pi}\int{d^3{\bi x}'}{1\over|{\bi x}-{\bi x}'|}
\left(
\begin{array}{c}
\rho(t',{\bi x}')/\varepsilon_0\\
\mu_0{\bi J}(t',{\bi x}')
\end{array}
\right),
\label{exact}
\ee
with $t'=t-|{\bi x}-{\bi x}'|/c$. Self-consistency results from requiring that the current density be minus the component of the displacement current along the magnetic field lines, and that the charge density is determined by the divergence of the perpendicular component of the electric field. 

The solution of equation (\ref{exact}) requires boundary conditions at the star. As in the Deutsch model, the corotation field inside the star determines the electric field at the surface, and the self-consistent field must join onto the interior solution continuously across the surface. This implies that the plasma immediately above the surface is nearly corotating, and deviations from corotation  increase with height above the surface. 

To describe the effects of partial current screening in detail one needs to solve equation (\ref{exact}). The screening cancels only the parallel component of the displacement current, $\varepsilon_0\partial{\bi E}_{\rm ind}/\partial t$, and we are interested in the resulting modified inductive field. 

One can solve equation (\ref{exact}) iteratively for sufficiently small $\sin\alpha$.  Identifying the zeroth order solution as the unscreened field due to the rotating dipole in vacuo, the current density is identified as the parallel component of ${\bi J}_{\rm sc}$, given by equation (\ref{scd2}), and the current in equation (\ref{exact}) is identified as this current directed along the dipolar field lines. The charge density is determined by integrating $\partial\rho/\partial t=\partial J_{{\rm sc}\parallel}/\partial s$, where $s$ denotes distance along the dipolar field line, with respect to time.  With these terms on the right hand side of equation (\ref{exact}), the solution is used to calculate the first order correction to the electric field and the associated current and charge densities. The second order corrections are found by repeating the calculation with the first order charge and current densities. To lowest order, the inductively induced drift modifies the rigid corotation that applies to zeroth order.

Hence the inductive electric field is given by,

\begin{equation}
{\bi E}_{\rm ind}(t,{\bi x}) = {\bi E}_{\rm ind}^{(0)}(t,{\bi x}) + {\bi E}_{\rm ind}^{(1)}(t,{\bi x}) + ... + {\bi E}_{\rm ind}^{(k)}(t,{\bi x}) + {\bi E}_{\rm ind}^{(k+1)}(t,{\bi x}) + ...,
\end{equation}
\\where in the Lorentz gauge, ${\bi E}_{\rm ind}^{(k+1)}(t,{\bi x})$ is a function of ${\bi E}_{\rm ind}^{(k)}(t,{\bi x})$, which can be determined from $\phi^{(k)}(t,{\bi x})$ and ${\bi A}^{(k)}(t,{\bi x})$ with,

\be
\left(
\begin{array}{c}
\phi^{(k+1)}(t,{\bi x})\\
{\bi A}^{(k+1)}(t,{\bi x})
\end{array}
\right)={1\over4\pi}\int{d^3{\bi x}'}{1\over|{\bi x}-{\bi x}'|}
\left(
\begin{array}{c}
\rho^{(k)}(t',{\bi x}')/\varepsilon_0\\
\mu_0{\bi J}^{(k)}(t',{\bi x}')
\end{array}
\right),
\label{iteration}
\ee
where ${\bi E}_{\rm ind}^{(0)}(t,{\bi x})$ represents the unscreened field due to the rotating dipole in vacuo.

The model used in section \ref{sect:nonrigid} corresponds to the first order term in the iteration in $\sin\alpha\ll1$. The iteration procedure must converge rapidly if $\sin\alpha$ is sufficiently small, but there  is no simple way of determining the range of validity of the approximation to first order in $\sin\alpha$.


\begin{thebibliography}{}

\bibitem[\protect\astroncite{Arons}{2004}]{Arons2004PWN}
Arons, J.: 2004,
\newblock {\em Advances in Space Research} {\bf 33}, 466

\bibitem[\protect\astroncite{Beloborogov \& Thompson}{2007}]{BT07}
Beloborogov, A.~M. \& Thompson, C.: 2007,
\newblock {\em ApJ} {\bf 657}, 967

\bibitem[\protect\astroncite{Beskin
  et~al.}{1993}]{PhysicsOfThePulsarMagnetosphere}
Beskin, V., Gurevich, A.~V., \& N., I.~Y.: 1993,
\newblock {\em Physics of the Pulsar Magnetosphere},
\newblock CUP

\bibitem[\protect\astroncite{Bogovalov}{1999}]{Bogovalov1999}
Bogovalov, S. V.: 1999,
\newblock {\em Astron. Astrophys.} {\bf 349}, 1017

\bibitem[\protect\astroncite{Cheng et~al.}{1986}]{OuterGap1986a}
Cheng, K.~S., Ho, C., \& Ruderman, M.: 1986,
\newblock {\em ApJ} {\bf 300}, 500

\bibitem[\protect\astroncite{Contopoulos}{2005}]{Contopoulos2005} 
Contopoulos, I.: 2005,
\newblock {\em A\&A} {\bf 442}, 579

\bibitem[\protect\astroncite{Deutsch}{1955}]{Deutsch1955}
Deutsch, A.~J.: 1955,
\newblock {\em Annales D'Astrophysique} {\bf 18}, 1

\bibitem[\protect\astroncite{Filippenko \&
  Radhakrishnan}{1982}]{FilippenkoRadhakrishnan1982}
Filippenko, A.~V. \& Radhakrishnan, V.: 1982,
\newblock {\em ApJ} {\bf 263}, 828

\bibitem[\protect\astroncite{Goldreich \& Julian}{1969}]{Goldreich1969}
Goldreich, P. \& Julian, W.~H.: 1969,
\newblock {\em ApJ} {\bf 157}, 869

\bibitem[\protect\astroncite{Gunn \& Ostriker}{1969}]{GO69}
Gunn, J.~E. \& Ostriker, J.~P.: 1969,
\newblock {\em Phys. Rev. Lett.} {\bf 22}, 728

\bibitem[\protect\astroncite{Hinotori et~al.}{2003}]{Hetal03}
Hinotori, K., Harding, A.~K., \& Shibata, S.: 2003,
\newblock {\em ApJ} {\bf 591}, 334

\bibitem[\protect\astroncite{Kirk et~al.}{2007}]{KirkLyubarskyPetri07}
Kirk, J. G., Lyubarsky, Y., \& P$\acute{\rm e}$tri, J.: 2007,
\newblock {\em arXiv:astro-ph/0703116v2}

\bibitem[\protect\astroncite{Kramer et~al.}{2006}]{Ketal06}
Kramer, M., Lyne, A.~G., O'Brien, J.~T., Jordan, C.~A., \& Lorimer, D.~R.:
  2006,
\newblock {\em Science} {\bf 312}, 549

\bibitem[\protect\astroncite{Levinson et~al.}{2005}]{Letal05}
Levinson, A., Melrose, D., Judge, A., \& Luo, Q.: 2005,
\newblock {\em ApJ} {\bf 631}, 456

\bibitem[\protect\astroncite{Lyne et~al.}{2010}]{LH}
Lyne, A., Hobbs, G., Kramer, M., Stairs, I., \& Stappers, B.: 2010,
\newblock {\em Science} {\bf 329}, 408

\bibitem[\protect\astroncite{Lyne \& Manchester}{1988}]{LyneManchester1988}
Lyne, A.~G. \& Manchester, R.~N.: 1988,
\newblock {\em MNRAS} {\bf 234}, 477

\bibitem[\protect\astroncite{Melatos \& Melrose}{1996}]{MM96}
Melatos, A. \& Melrose, D.~B.: 1996,
\newblock {\em MNRAS} {\bf 279}, 1168

\bibitem[\protect\astroncite{Melrose}{1967}]{Melrose1967}
Melrose, D.~B.: 1967,
\newblock {\em Planet. Space. Sci.} {\bf 15}, 381

\bibitem[\protect\astroncite{Melrose \& Luo}{2004}]{ML04}
Melrose, D.~B. \& Luo, Q.: 2004,
\newblock {\em MNRAS} {\bf 352}, 915

\bibitem[\protect\astroncite{Michel}{2004}]{M04}
Michel, F.~C.: 2004,
\newblock {\em Adv. Space Phys.} {\bf 33}, 542

\bibitem[\protect\astroncite{Rees \&
  Gunn}{1975}]{ReesGunn1974}
Rees, M. J. \& Gunn, J.~E.: 1974,
\newblock {\em MNRAS} {\bf 167}, 1

\bibitem[\protect\astroncite{Ruderman \&
  Sutherland}{1975}]{RudermanSutherland1975}
Ruderman, M. \& Sutherland, P.~G.: 1975,
\newblock {\em ApJ} {\bf 196}, 51

\bibitem[\protect\astroncite{Scharlemann et~al}{1978}]{SAF78}
Scharlemann, E. T., Arons, J. \& Fawley, W. M.: 1978,
\newblock {\em ApJ} {\bf 222}, 297

\bibitem[\protect\astroncite{Sang \& Lysak}{2006}]{SL06}
Sang, Y. \& Lysak, R.~L.: 2006,
\newblock {\em PRL} {\bf 96}, 145002

\bibitem[\protect\astroncite{Sturrock}{1971}]{Sturrock1971}
Sturrock, P.: 1971,
\newblock {\em ApJ} {\bf 164}, 529

\bibitem[\protect\astroncite{Takata et~al.}{2010}]{Tetal10}
Takata, J., Wang, Y., \& Cheng, C.~S.: 2010,
\newblock {\em ApJ} {\bf 726}, 44

\bibitem[\protect\astroncite{Timokhin}{2010}]{T10}
Timokhin, A.~N.: 2010,
\newblock {\em MNRAS} {\bf 408}, 2092

\bibitem[\protect\astroncite{Usov}{1994}]{U94}
Usov, V.~V.: 1994,
\newblock {\em MNRAS} {\bf 267}, 1035

\bibitem[\protect\astroncite{Weltevrede et~al.}{2006}]{Wetal06}
Weltevrede, P., Edwards, R.~T., \& Stappers, B.: 2006,
\newblock {\em A\&A} {\bf 445}, 243

\bibitem[\protect\astroncite{Weltevrede et~al.}{2007}]{Wetal07}
Weltevrede, P., Stappers, B.~W., \& Edwards, R.~T.: 2007,
\newblock {\em A\&A} {\bf 469}, 607

\end{thebibliography}
\end{document}